\documentclass[proof]{WileyASNA-v2}

\articletype{Article Type}%

\newcommand{\slsh}[1]{\not \! #1}

\received{2023}
\revised{2023}
\accepted{2023}

\begin{document}

\title{Particle interaction strengths modified by magnetic fields}

\author[1,2]{Gabriella Piccinelli*}

\author[2]{Angel S\'anchez}

\author[2]{Jorge Jaber-Urquiza}

\authormark{Gabriella Piccinelli}

\address[1]{\orgdiv{Centro Tecnol\'ogico, Facultad de Estudios Superiores Arag\'on}, \orgname{Universidad Nacional Aut\'onoma de M\'exico (UNAM)}, \orgaddress{\state{Estado de M\'exico}, \country{M\'exico}}}

\address[2]{\orgdiv{Facultad de Ciencias}, \orgname{Universidad Nacional Aut\'onoma de M\'exico (UNAM)}, \orgaddress{\state{Ciudad de M\'exico}, \country{M\'exico}}}

\corres{*Av. Rancho Seco S/N, Col. Impulsora Popular Av\'icola, Nezahualc\'oyotl - Estado de M\'exico, 57130.\email{itzamna@unam.mx}}

\abstract{\textbf{ABSTRACT}\\

Magnetic fields are everywhere in the Universe and in our everyday life and many processes are affected by their presence, generating a rich phenomenology that depends also on other possible external agents.
We review here some results, both from our workgroup and from other research groups,  about the effect of magnetic fields on particles interaction processes, focusing mainly on recent results, but without losing sight on early seminal works on this topic. A vast assortment of physical situations and of analytical and numerical approaches can be found in the literature in this subject, making the comparison between them not straightforward. Our aim is to focus attention on differences and similarities between the different situations and approaches, looking for a systematization scheme that could be predictive, once the role played by each physical ingredient could be understood. The main purpose of this work is to find some physical explanations of the ongoing processes.
 }

\maketitle

\section{Introduction}

Magnetic fields are pervasive in our everyday life, in our laboratories and in the Universe.

In the relativistic heavy ions collisions, that take place in the RHIC (Relativistic Heavy Ion Collider) and LHC (Large Hadron Collider) facilities, a magnetic field presence is theorized in the initial stages and can reach enormous intensities of the order of $10^{18}$ Gauss. This leads to a broad number of studies on the subject and the discovery of novel effects that can be tested in this kind of collisions.

Cosmic magnetic fields are present at all scales, being their origin presently unclear: they could either be primordial -where by that we use to mean before the recombination epoch- or produced during the structure formation process. Their presence at the largest scales~\cite{Acciari_2023}  and in almost empty regions \cite{Neronov_2010, Tavecchio2010,Dolag_2010} strongly supports their  primordial origin. If this is the case, their contribution  in cosmological events can be fundamental. 

In both these areas, where fundamental particles, some of them charged, experience extreme conditions, the presence of magnetic fields gives rise to an entire area of study, that has to be continuously explored. We are interested here, in particular, in their effect on the interaction of particles.
In this field there is a vast literature: their influence over interaction processes has been studied in many contexts, in the presence of different physical ingredients, and with a variety of approaches.  Nonetheless, a physical clear explanation of the ongoing process and the obtained results is barely presented. Besides, comparison between different results is not straightforward due to the assortment of physical components and  different kinematic regimes.

The calculations necessary for the study of this topic typically require of an approximation scheme, depending on the method employed, the physical conditions present in the different situations and the hierarchy of the involved scales. Along the development of the subject, there has been some proposals of approximation procedures, for different regimes, that have afterwards been applied to different situations. From this derives the importance of having a good control on the different specific regimes that can be of interest. 
Some possible important ingredients are: strong or weak magnetic fields, finite or zero
temperature, high or low particle momenta, spin and charge of the particles involved in the interaction. 

In the literature, the interaction process is favored in some cases (\cite{Bali2018}; \cite{Bandyopadhyay}; \cite{Bhattacharya}; \cite{ErdasLissia2003}; \cite{KuznetsovPLB1998,KuznetsovPRD2006}; \cite{Mikheev2001}; \cite{Satunin2015}; \cite{TSAI.1,TSAI.2}; \cite{Urrutia1978}).
In some others it is inhibited (\cite{Bandyopadhyay2018}; \cite{Kawaguchi2017}; \cite{Sogut2017}) or it can transit from one behavior to the other, depending on the kinematic regime of the progenitor particle (\cite{Chistyakov1998}; \cite{Ghosh2017}; \cite{Jaber}; \cite{GhoshChandra2019}).

Facing this situation, the aim of our work is to put together some results that can be found in the literature, both from our work team  and other research groups, highlighting differences and similarities between the different situations, trying to find some systematization scheme that could guide us through the study of this field of knowledge.  We concentrate on a few early seminal works and on the most recent ones. No need to say that, since it is a very active research area, our review is far to be exhaustive.

We have organized our work as follows: after a brief presentation in section~\ref{sec2} of the formalisms that can be employed for introducing the magnetic field effect in the particles propagators, we review in section~\ref{sec3} some applications that can be found in the literature, dividing it in three subsections, corresponding to the hierarchy of the magnetic field strength with respect to other relevant scales: strong field approximation, weak field approach and  general methodologies that allow to study different regimes. We end presenting some preliminary conclusions about the systematization scheme we are willing to build.

\section{Methods for including the effect of magnetic fields on particles propagation}\label{sec2}

The magnetic field effect is introduced in particle processes that involve charged particles through two different methods: Schwinger's proper time~\cite{Schwinger1951} and Ritus' eigenfunctions~\cite{Ritus:1978cj}. In the former method, any charged particle propagator dressed with the magnetic field, in general, has the form~\cite{Erdas2009}
\begin{eqnarray}
     G(x,y)=\Omega(x,y)\int\frac{d^4p}{(2\pi)^4}e^{-ip\cdot (x-y)}\tilde{G}(p),    
\end{eqnarray}
where 
\begin{equation}
    \Omega(x,y)\equiv \exp\left(-iq\int_x^y d\zeta^\mu A_\mu(\zeta)\right)
\end{equation}
is the Schwinger's phase, $q<0$ the particle charge, and $\tilde{G}(p)$ is a translational invariant part whose explicit form depends on the particle's spin.

In the case of a constant magnetic field along $z$-direction, the explicit form of $\tilde{G}(p)$ can be decomposed as follows
\begin{equation}
\tilde{G}(p)= \int_0^\infty \frac{d\tau}{\cos(qB\tau)} e^{i \tau(p_{||}^2-m^2)}e^{i\tau p_{\perp}^2\frac{\tan(qB\tau)}{qB\tau}} \text{Spin}(p)  
\end{equation}
with $\tau$ the Schwinger's proper time and $\text{Spin}(p)$ a structure that encodes the particle's spin interaction with the magnetic field. Note that the magnetic field splits the particle's momentum into parallel and perpendicular components $p_{||}^2\equiv p_0^2-p_z^2$ and $p_{\perp}^2=-p_x^2-p_y^2$, with respect to the magnetic field direction.

The spin factor for a charged scalar, becomes
\begin{equation}
    \text{Spin}(p)=1.
\end{equation}
Meanwhile, for a charged fermion, it has the form
\begin{equation}
    \text{Spin}(p)=(m +\slsh{p}_{||})e^{i qB\tau \Sigma_3}+\frac{\slsh{p}_\perp}{\cos(qB\tau)},
\end{equation}
where  $p\hspace{-.5em}/ =\gamma^\mu p_\mu$, with $\gamma^\mu$ the Dirac matrices, and $\Sigma_3=i\gamma^1\gamma^2$ the spin matrix along $z$. 

Finally, for charged vector, it reads  
\begin{equation}
    \text{Spin}(p)=g_{||}^{\mu\nu}+g_{\perp}^{\mu\nu}\cos(2qB\tau)+\hat{F}^{\mu\nu}\sin(2qB\tau),
\end{equation}
where $g^{\mu\nu}$ the metric tensor, which satisfies
\begin{equation}
    g^{\mu\nu}=  g_{||}^{\mu\nu}+  g_{\perp}^{\mu\nu},
\end{equation}
and $\hat{F}^{\mu\nu}\equiv F^{\mu\nu}/B$ the field strength tensor.

In the Ritus' method the charged particle propagators are represented in term of the so called Ritus' eigenfunctions. These eigenfunctions, obtained by solving the equation of motion of charged particles in presence of an external magnetic field, are used to represent the charged particle propagators as
\begin{equation}
    G(x,y)=\sum_{n=0}^\infty \int\frac{d^3p}{(2\pi)^3}\frac{\mathbf{E}_p(x)\ \text{Spin}(\bar{p})\ \overline{\mathbf{E}}_p(y)}{\bar{p}^2-m^2}
\end{equation}
where, in the asymmetric gauge $A^\mu(x)=(0,0,Bx^1,0)$, the Ritus eigenfunctions are given by 
\begin{equation}
    \begin{split}
        \mathbf{E}_{p}&\equiv \sum_{\sigma}N(n)e^{i(p_0 x_0-p_2 x_2 - p_3 x_3)}D_n(\rho){\bf F}_{\sigma},\\
        \rho&\equiv \sqrt{2|qB|}\left(x_1-\dfrac{p_2}{qB}\right),\\
        N(n)&\equiv\dfrac{\left(4\pi|qB|\right)^{1/4}}{\sqrt{n!}},
    \end{split}
\end{equation}
with $D_n(\rho)$ the parabolic cylinder functions of order $n$, and ${\bf F}_{\sigma}$ a spin structure that depends on the kind of particle one works with. The summation over $\sigma$ accounts for the different spin projections along the magnetic field. Note that in this formalism the discrete number $n$, identified with the Landau level, rises as a consequence of the confinement along the transverse direction respect to the magnetic field.

In the fermionic case \cite{RITUS} 
\begin{eqnarray}
    \text{Spin}(\bar{p})&\equiv& \slsh{\bar{p}}+m,\nonumber \\
    \bar{p}^\mu&\equiv& (p^0,0,sign{(qB)}\sqrt{2n|qB|},p^3),\nonumber \\
    {\bf F}_{\sigma}&\equiv&\frac{1}{2}\left(1+i \sigma sign(qB)
    \gamma^1\gamma^2\right),
    \nonumber \\
    n&\equiv&\frac{p_{||}^2-m^2}{2|qB|}+\dfrac{\sigma}{2}sign(qB)-\dfrac{1}{2},
\end{eqnarray}
with $sign(qB)$ the sign function.

In the scalar case \cite{Lawrie1997, Correa2013} 
\begin{eqnarray}
    \text{Spin}(\bar{p})&\equiv& 1,\nonumber \\
    \bar{p}^\mu&\equiv& (p^0,0,sign{(qB)}\sqrt{(2n+1)|qB|},p^3),\nonumber \\
    {\bf F}_{\sigma}&\equiv& 1,
    \nonumber \\
    n&\equiv&\frac{p_{||}^2-m^2}{2|qB|}-\dfrac{1}{2}.
\end{eqnarray}

Finally, for a massive charged vector field in the Feynman gauge, we have
\begin{eqnarray}
    \text{Spin}(\bar{p})&\equiv& g^{\mu\nu},\nonumber \\
    \bar{p}^\mu&\equiv& (p^0,0,sign{(qB)}\sqrt{(2n+2\sigma+1)|qB|},p^3),\nonumber \\
    {\bf F}_{\sigma}&\equiv& \Gamma^{\alpha\beta}(\bar{p}),
    \nonumber \\
    n&\equiv&\frac{p_{||}^2-m^2}{2|qB|}+\sigma sign(qB)-\dfrac{1}{2},
\end{eqnarray}
for the explicit form of $\Gamma^{\alpha\beta}(\bar{p})$ see \cite{Elizalde2002}.

\section{Applications}\label{sec3}

With the formalisms of the previous section, we have the mechanisms for dressing the charged particles propagators with the magnetic field effects for studying any particles process. We will report here some study cases 
of magnetized particles interactions.
Since the presented methods generally require some approximation process, we divided the cases that we are reporting following the hierarchy of the involved scales: strong magnetic field; weak magnetic field with two possibilities for the kinematic regime: small and large momenta of the progenitor particle; general approaches for arbitrary magnetic fields strengths.

\subsection{Strong magnetic field approximation}
 
In the case that the interest falls on the r\^ole played by a magnetic field in a warm inflation scenario in the early universe (\cite{Piccinelli2022}), several aspects have to be taken into account. One of them is the effect on the decay process of the inflaton to light fields, possibly through the intermediation of heavy boson and fermion fields. Indeed, a successful warm inflation model (\cite{BERERA2003};\cite{HallMoss}) has been embedded in a supersymmetric scenario with a two step decay process: $\Phi \rightarrow \chi \rightarrow y$, where $\Phi$ is the inflaton field, $\chi$ represents the heavy sector and $y$ the light one, being that these two ingredients together ensure that the interaction of the inflaton with other fields during inflation does not spoil the flatness of the effective potential. The heavy and light fields can be charged and feel the magnetic field influence.
In this framework, if the heavy sector feels the presence of the magnetic field, then this physical quantity represents the dominant scale for the light sector. Thereby, working in a strong magnetic field regime and in the low transverse momentum approximation, in (\cite{Piccinelli2022}), the decay probability of a scalar field to two fermions was estimated, making use of Schwinger proper time method and of the optical theorem which relates the imaginary part of the self-energy with the decay width:
\begin{equation}
    \Gamma(p)\equiv-\frac{Im(\Sigma(P))}{2E},
\end{equation}
 with $\Sigma$ the scalar self-energy and $E$ its energy. It was found to grow linearly with $qB$, with $q$ a representative charge of the particles involved in the interaction and $B$ the magnetic field strength. Although this represents only one channel of decay of the charged particles involved, it seems to reinforce one required condition for warm inflation, which is that the inflaton interaction with other fields becomes important during the inflationary stage.

Note that, although the decay was analyzed in the lowest Landau level (LLL), the effective potential for the heavy sector that modifies the inflaton potential was computed for an arbitrary magnetic field strength, which implicitly gives us information about the magnetic field effect on the ground state where other particles move, affecting their masses and interactions.

The encountered linear $qB$ dependence came from a Gaussian integration over transverse momentum, where the factor $\exp{\left(-p_{\perp}^2/{qB}\right)}$, which appears in the charged propagators written in terms of the Landau Levels \cite{GusyninMiranskyShovkovy},   acts as a statistical weight for the momenta distribution, confining the fermions in a region $r \sim (qB)^{-1/2}$. In this process, momentum conservation in the transverse direction was linked with the magnetic field scale and the standard self-energy  dimensions (in this case, energy squared) have to be invoked.

A similar dependence was established in (\cite{MIRANSKY20151}), when examining the decay of a Nambu-Goldstone boson to two quarks in the LLL (see Eq.~(273)).

\cite{Chino1Yang}, working in a two-flavor Nambu-Jona-Lasinio (NJL) model (for a review see~\cite{Buballa2005}), study the effect of different external agents on the properties of a neutral boson, such as the mass, the coupling constant and the decay width. We focus here on the part of their work that considered the effects of an strong external magnetic field on the neutral boson decay process, which emerge from the magnetization of the quarks that constitute the boson. In particular, they study the decay $\sigma \rightarrow \pi_0 \pi_0$ in presence of an homogeneous magnetic field, by computing its corresponding triangular diagram, using the Ritus approach to write the magnetic field dressed propagators. Their numerical results show that the magnetic field enhances the amplitude of the vertex and, consequently, the decay rate.

 \subsection{Weak magnetic field approximation}
 
\begin{figure}
    \centering
    \includegraphics[width=6cm]{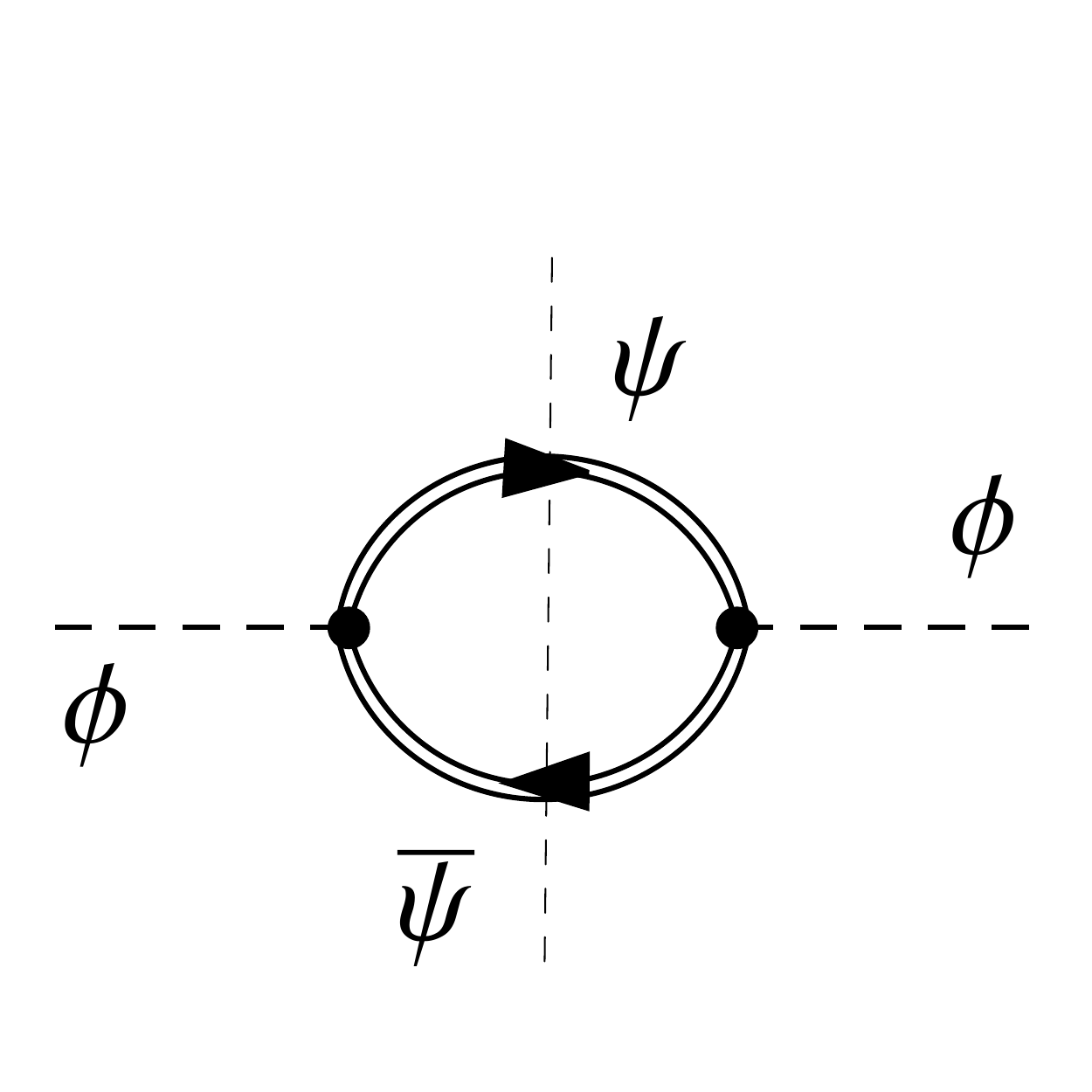}
    \caption{Diagrammatic description of the Cutosky rule applied to the scalar self-energy in interaction with a couple of fermions (\cite{Jaber}).}
\label{figCutosky}
\end{figure}

 For a general decay process in presence of an external magnetic field, there are different aspects that could be relevant in the behavior of the decay width that have to be taken into account, for example, the kinematics of the particle plays an important r\^ole in the decay process and had to be carefully considered in the expressions obtained by~\cite{TSAI.1,TSAI.2}. Another ingredient that could be relevant is the spin. In particular, the studies done in~\cite{PiccinelliSanchezEscalar,Jaber} explore different spins of the daughters particles by analyzing the decay of a neutral scalar particle into charged scalars ($\Phi\longrightarrow\phi\phi^*$) and charged fermions ($\Phi\longrightarrow\bar{\psi}\psi$), respectively, in presence of an uniform magnetic field. Both works consider the weak magnetic field limit (compared with the decay product masses $qB\ll m^2$, with $q$ its electric charge) for different kinematic regions of the neutral scalar particle. The methodology used is the following: the one-loop self-energy of the neutral scalar particle is computed and then the decay width is obtained by applying the optical theorem
 and the Cutkosky rules, illustrated in Fig.\ref{figCutosky}. As the particles inside the loop are charged, the Schwinger proper time formalism is used to write the full propagators dressed with the magnetic field. In \cite{PiccinelliSanchezEscalar}, the effect of a thermal bath was also included through the Matsubara frequencies in the imaginary time formalism. The pair creation process was found to be inhibited, meanwhile temperature had the opposite effect. 
 
 These results were obtained through a perturbative approach for weak fields and then the computation in vacuum was reproduced following another approximation method, extensively used in literature (see e.g. \cite{TSAI.1}), dubbed crossed approximation in the sense that the magnetic field is weak, although the transverse momentum is allowed to be large. 
 The suppression due to the magnetic field on the scalar decay was still present, and even more pronounced. The authors offered a possible physical explanation for this  difference referring to the fact that the crossed approximation can be related with an infinite photon insertion, meanwhile the perturbative approximation takes into account only two photons, in agreement with Furry’s theorem.
 
The study of other processes, as $\nu \rightarrow W^{+} + e^{-}$ \cite{KuznetsovPLB1998,KuznetsovPRD2006,ErdasLissia2003,Bhattacharya} and $\gamma \rightarrow e^{+} + e^{-}$ \cite{TSAI.1,TSAI.2,Urrutia1978}, with the crossed field approximation, led to the opposite behavior: an enhancement of pair creation in presence of magnetic fields. In \cite{PiccinelliSanchezEscalar}, the difference was suggested to be related to the analytical structure of the self-energy dictated by the spin of the particles involved in the processes, generating  linear  and quadratic divergences for $\nu \rightarrow W^{+} + e^{-}$ and $\gamma \rightarrow e^{+} + e^{-}$, respectively, and a logarithmic one in the scalar case. After all, different dependencies on the momentum in the loop are expected to determine the response to the external magnetic field.

\begin{figure}
    \centering
    \includegraphics[width=8cm]{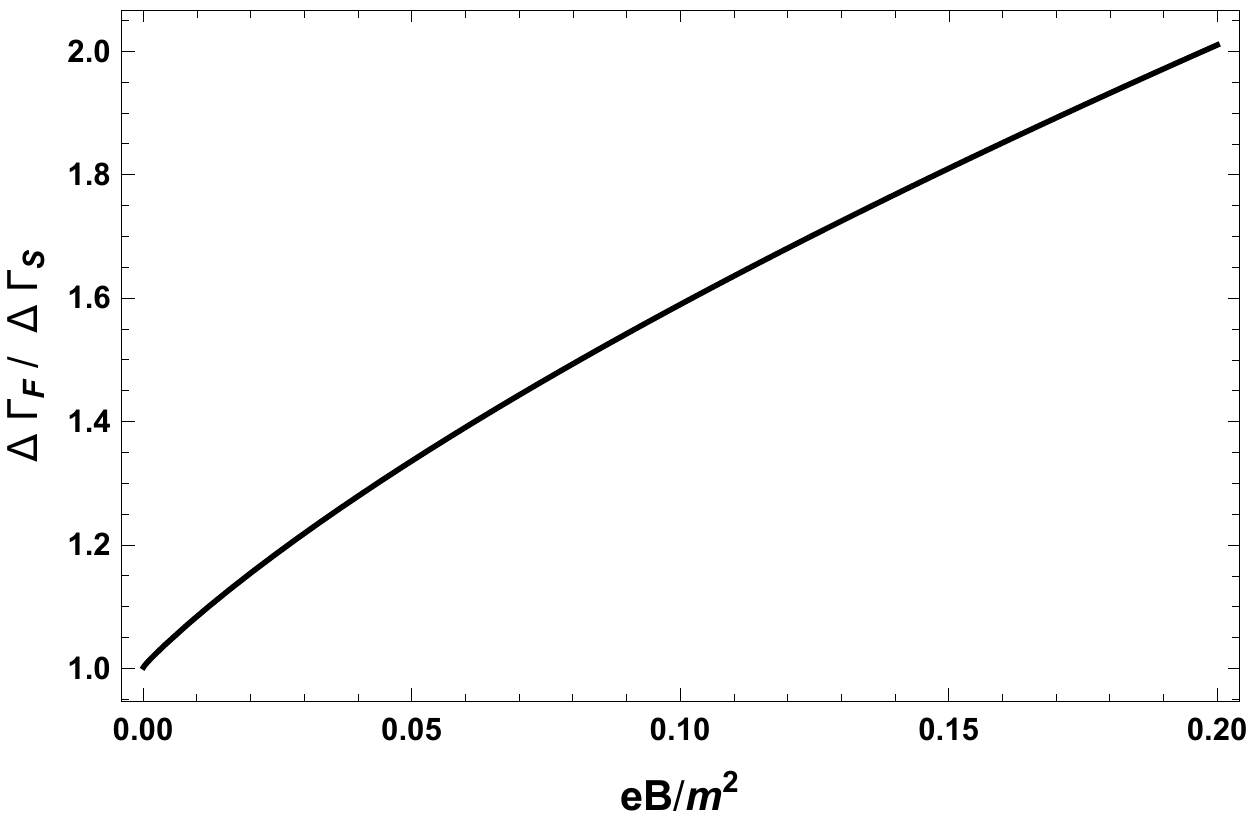}
    \caption{ Ratio of the decay responses to the external magnetic field for two different channels, $\phi\rightarrow \psi+\psi$ and $\phi\rightarrow \varphi+\varphi$, where $\psi$ and $\varphi$  are charged  fermions and scalars, respectively. We plot this ratio as a function of the magnetic field in the high momentum approximation, for $p_\perp/m=10^3$, and  $p/m=5$.}
    \label{figJPS}
\end{figure}

In ~\cite{Jaber}, the behavior of the decay width to charged fermions was shown to depend on the kinematics of the progenitor particle: suppressed by the magnetic field in the low transverse momentum regime and enhanced in the high transverse momentum regime. Besides, in order to shed some light on the r\^ole of the loop particles spin, a comparison between the results of both works for the weak field limit with high transverse momentum regime was presented and seemed to indicate that the magnetic field impact is more important on the decay process to charged fermions than to charged scalars, as can be seen in Fig.~\ref{figJPS}. These results suggest that the spin-magnetic field interaction increases  the progenitor
scalar particle decay into a pair of charged fermions as compared to a pair of charged scalars.

In Fig.~\ref{figJPS}, the decay response is defined as the ratio between the decay width in presence of an external magnetic field and vacuum:
\begin{equation}
\Delta\Gamma(p,p_\perp,m,eB)\equiv\dfrac{\Gamma^{^B}(p,p_\perp,m,eB)}{\Gamma_{\text{\textit{vac}}}(p,p_\perp,m)}\hspace{1mm}.
\label{eq.compara-DeltaGamma}
\end{equation}

At this point, some remarks about the different approximation approaches are in order. In \cite{PiccinelliSanchezEscalar} and \cite{Jaber} an exact expression for the neutral scalar self-energies in presence of the magnetic field, up to Schwinger proper time integrals, is presented, writing the charged particles propagators in their complete form (without any approximations). Then, two different weak magnetic field approximation are taken by imposing hierarchies among the three energy scales: the magnetic field, the decay product masses and the progenitor particle perpendicular momentum with respect to the magnetic field direction. The weak magnetic field with low transverse momentum is obtained when the hierarchy is the following $qB\ll m^2$ and $|p_\perp|\lesssim m$, here, a power series expansion on $eB$ can be safely performed in the whole expression for the decay width. 
On the other hand, for the hierarchy $qB\ll m^2\ll |p_\perp|^2$, which defines the weak magnetic field with high transverse momentum,  the power series expansion on $qB$ needs caution since there are terms that mix these three energy scales in such a way that the expansion is no longer valid.

All of the above can be easily seen in the scalar self-energy in the presence of an external  magnetic field, given by  
\begin{eqnarray}
     \Pi_B(p)&=&\frac{g^2}{32\pi^2}\int_0^\infty ds\int_{-1}^1dv \frac{qB}{\sin(qBs)}
          e^{-i s m^2} \nonumber \\
          &&\hspace{4em}\times\
          e^{is\left[\frac{1}{4}(1-v^2)p_{||}^2-p_{\perp}^2\frac{\cos(qBsv)-\cos(qBs)}{2qBs\sin(qBs)}\right]}. 
\end{eqnarray}
Note that the crossed field approximation consists in expanding the argument of the exponential up to $(qBs)^2$, with $s$ the Schwinger proper time,  but not the exponential itself, since it contains a factor $p_\perp^2$. The expansion in $qBs$ can be done on the rest of terms by considering that the main contribution to the integral over $s$ comes from the region $qBs \ll 1$.

In contrast, a different way to obtain the weak field limit is to expand the charged particles propagators dressed with the magnetic field effects in power series on $eB$ at the beginning of the calculation (before computing the loop momentum integrals). This approach is commonly used in literature~\cite{TaiwanDebil,AyalaDebilGluon,GluonDebil} without any assumption on the particles kinematics. 
This point is addressed in \cite{Jaber}, where their calculations allow to explore both kinematics regimes, making more transparent how, in the commonly used approach, the low momentum regime is implicitly assumed.
In this sense, the methodology that begins from exact expression is more general because it allows to analyze different magnetic field regions and kinematics regimes from an analytical expression valid for an arbitrary magnetic field strength.

 \subsection{Arbitrary magnetic field strength}

 \noindent The decay channel of the Higgs boson into two photons has played an important role in particle physics, since it was the channel through which the Higgs was first detected back in 2012~\cite{ATLASdetect,CMSdetect,Confirma1,Confirma2,Confirma3}. Nowadays, Higgs physics is in the precision era, where the experimental and theoretical sides are working to determine its properties with extreme accuracy. Then, corrections due to external agents could be considered in the theoretical calculations, since their presence may modify the Higgs properties. We will study here the general properties of the process in presence of an external magnetic field.

\begin{figure}[h]
	\centering
	\includegraphics[width=6.5cm]{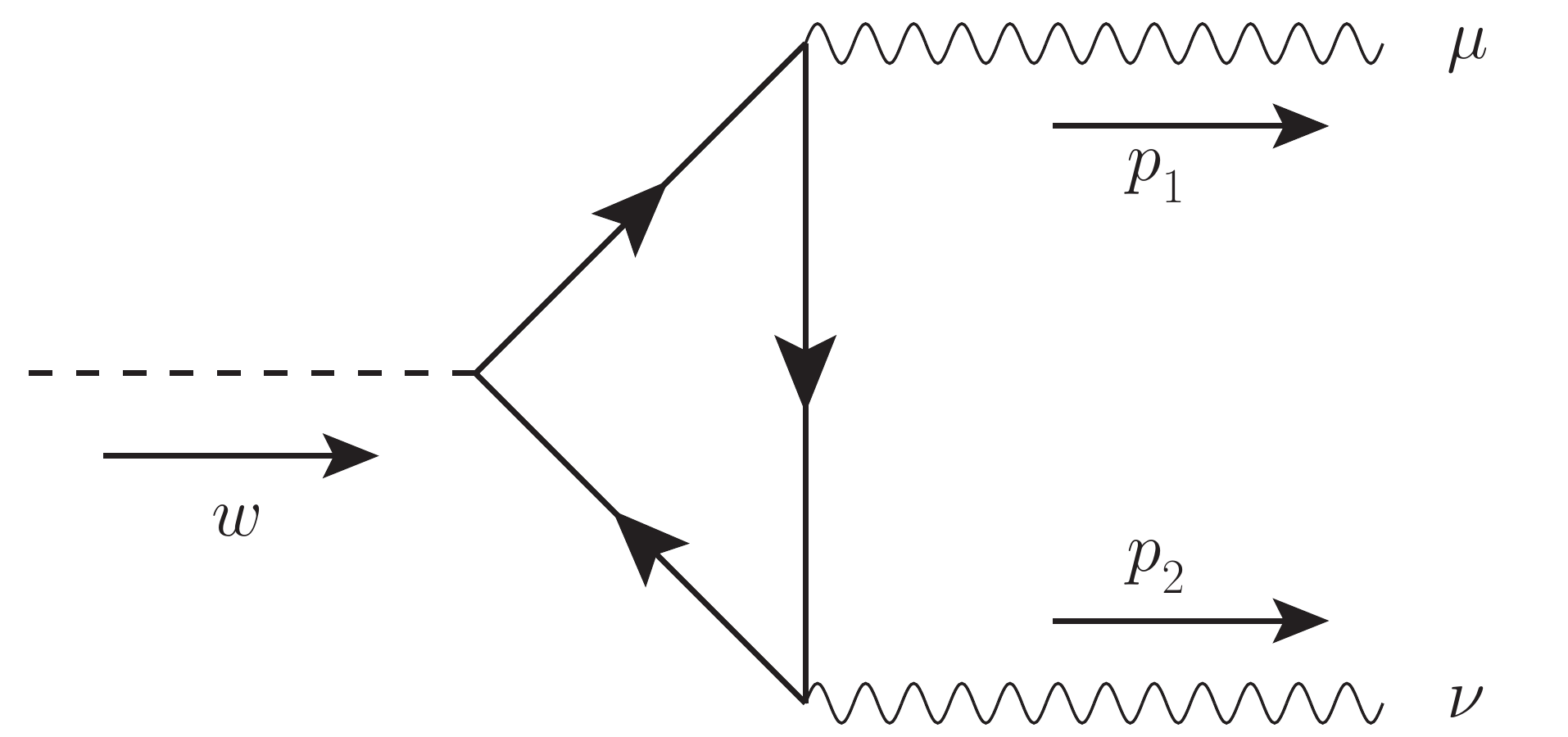} %
	\caption{One-loop Feynman diagram that accounts for the Higgs boson decay into two photons.}
	\label{figuratriagulo}
\end{figure}

The Higgs boson decay to two photons
\begin{equation*}
	H\longrightarrow\gamma^\mu+\gamma^\nu
\end{equation*}
is described by the following invariant matrix element
\begin{equation}                                   
        \begin{split}
            \mathcal{M}=&\big<(p_1,\mu,\lambda_1),(p_2,\nu,\lambda_2)|w\big>\epsilon_\mu^*\left(p_1,\lambda_1\right)\epsilon_\nu^*\left(p_2,\lambda_2\right)\\
    \equiv&\mathcal{M}^{\mu\nu}\epsilon_\mu^*\left(p_1,\lambda_1\right)\epsilon_\nu^*\left(p_2,\lambda_2\right),
        \end{split}
    \label{eq.vacio-amplitud}
\end{equation}
where $p_i$ and $\lambda_i$, with $i=1,2$, are the momenta and polarization states of the out-coming photons and $w$ is the Higgs boson momentum. $\mathcal{M}^{\mu\nu}$ encodes the overlap between the initial and final states. In the Standard Model there is not a direct interaction among these three particles, the leading order contributions to $\mathcal{M}^{\mu\nu}$ coming from radiative corrections associated to Feynman diagrams of the type shown in Fig.\ref{figuratriagulo}.

Based on general grounds, as symmetries and particle properties, we present the ideas behind the structure of the amplitude tensor and build it explicitly in the presence of an external magnetic field in a simple way.


\noindent The amplitude shown in Fig.~\ref{figuratriagulo} must, in general, satisfy the Ward identities~\cite{Ward,Taka}, that arise from the current conservation and assure the amplitude transversality:
\begin{equation}
	\label{eq.sinB-Slavnov1}
	p_1^\mu\mathcal{M}_{\mu\nu}(p_1,p_2)=0\mbox{\  \ and \ \ } 
	p_2^\nu\mathcal{M}_{\mu\nu}(p_1,p_2)=0.
\end{equation}
The amplitude must also be invariant to the photon exchange, namely:
\begin{align}
	\mathcal{M}^{\mu\nu}(p_1,p_2)=\mathcal{M}^{\nu\mu}(p_2,p_1).
	\label{eq.sinB-PropBosones}
\end{align}
Besides, the vertex must be invariant under the discrete transformations of charge conjugation $(C)$ and parity $(P)$~\cite{QEDlandau}, either in vacuum or in the presence of an external magnetic field. 

In what follows, we analyze its general tensor in the presence of an arbitrary constant magnetic field, described by the field strength tensor $F^{\mu\nu}$. In this situation, the number of independent four-vectors increases from two to eight~\cite{Batalin,RITUS,Hugo}, namely
\begin{equation}
	p_i^\mu,\ F^{\mu\nu}{p_i}_{\nu}, \ {F^\mu}_\alpha F^{\alpha\nu}{p_i}_{\nu} \ \text{  and  } \ F^{*\mu\nu}{p_i}_{\nu},
\label{newvectors}
\end{equation}
where $i=1,2$.
The dual electromagnetic field strength tensor
\begin{equation}
	F^*_{\mu\nu}\equiv\frac{1}{2}\epsilon_{\mu\nu\gamma\delta}F^{\gamma\delta},
\end{equation}
is considered to satisfy the cross field condition, $F^{\mu\nu}F^*_{\mu\nu}$=0.

In order to obtain orthogonal tensor structures, following~\cite{Batalin}, a set of orthogonal vectors can be built from the ones in Eq.~(\ref{newvectors}), as
\begin{equation}
\begin{split}
	l^{\mu}_i\equiv p_i^{\mu}, \  \
	L^{\mu}_i\equiv \hat{F}^{\mu\nu}{p_i}_{\nu},\  \
	L^{*\mu}_i\equiv \hat{F}^{*\mu\nu}{p_i}_{\nu}\ \\
    \mbox{ and } \
	G^{^\mu}_i\equiv \frac{l^2}{L^2}\hat{F}^{\mu\alpha}\hat{F}_{\alpha\beta}p_i^{\beta}+l_i^{\mu},
 \end{split}
\label{setspolvects}
\end{equation}
with $\hat{F}^{\mu\nu}\equiv F^{\mu\nu}/|B|$. In this way, the last three vectors in Eq.~(\ref{setspolvects}), can be used to describe the photon polarization states.

Then, given that each Lorentz index describes a photon with a given momentum, each set of vectors is related to a unique Lorentz index, so, the most general form of the amplitude, that fulfills Eq.~(\ref{eq.sinB-Slavnov1})~\cite{RitusPapan,RitusPapan2}, can be found:
\begin{equation}
	\begin{split}
		\mathcal{M}_{{qB}}^{\mu\nu}(p_1,p_2)=&
   a_1^{++}\hat{L}_{1}^{\mu}\hat{L}_{2}^{\nu}
  +a_2^{++}\hat{L}_{1}^{*\mu}\hat{L}_{2}^{*\nu}
  +a_3^{++}\hat{G}_{1}^{\mu}\hat{G}_{2}^{\nu}\\
  &+a_4^{+-}\hat{L}_{1}^{\mu}\hat{L}_{2}^{*\nu}+a_5^{+-}\hat{L}_{1}^{*\mu}\hat{L}_{2}^{\nu}
  +a_6^{-+}\hat{L}_{1}^{\mu}\hat{G}_{2}^{\nu}\\
  &+a_7^{-+}\hat{G}_{1}^{\mu}\hat{L}_{2}^{\nu}
  +a_8^{--}\hat{L}_{1}^{*\mu}\hat{G}_{2}^{\nu}
  +a_9^{--}\hat{G}_{1}^{\mu}\hat{L}_{2}^{*\nu},
	\end{split}
	\label{eq.MF.Esttensor1}
\end{equation}
where the superscripts ``$\pm$" in the  first and second position in the $a_j$ coefficients indicate their behavior under charge and parity transformations, respectively, and the ``hat" over vectors indicates normalization to unity. The subscript $qB$ indicates that the amplitude is dressed with the magnetic field. This structure is based on the electromagnetic tensor field parity and charge properties and obeys Furry's theorem~\cite{Batalin}.

Finally, considering on-shell photons (so that $G^{\mu} \propto l^\mu$) and by requiring photon exchange symmetry (Eq.~(\ref{eq.sinB-PropBosones})), the amplitude in presence of an external magnetic field becomes
\begin{equation}
	\begin{split}
	    \mathcal{M}_{{qB}}^{\mu\nu}(p_1,p_2)=&
       a_1^{++}\hat{L}_{1}^{\mu}\hat{L}_{2}^{\nu}
      +a_2^{++}\hat{L}_{1}^{*\mu}\hat{L}_{2}^{*\nu}\\
      &+a_4^{+-}\dfrac{1}{\sqrt{2}}\left(\hat{L}_{1}^{\mu}\hat{L}_{2}^{*\nu}
      +\hat{L}_{1}^{*\mu}\hat{L}_{2}^{\nu}\right).
	\end{split}
	\label{eq.MF.EsttensorOnShell}
\end{equation}

The above expression, that emerges from general grounds, is the form of the effective vertex $\mathcal{M}$ in terms of orthogonal tensor structures. Since $\mathcal{M}^2$ is related to the strength of the interactions between the fields, the structure in Eq.~(\ref{eq.MF.EsttensorOnShell}) allows us to quantify these intensities as the squares of the coefficients, being that each coefficient is related to the interaction of photons in specific polarization states.

Nonetheless, we do not known the explicit form of the amplitude which encodes the microphysics that allows the interaction between these three particles. In general, up to a certain order of approximation, the computation of $\mathcal{M}_{{qB}}^{\mu\nu}$, which requires a finite sum of Feynman diagrams, is not expressed in a closed form of orthogonal tensor structures as in Eq.~(\ref{eq.MF.EsttensorOnShell}) but, once computed, each coefficient can be obtained by projecting the whole vertex with its corresponding tensor structure, that is
\begin{equation}
    \begin{split}
	\label{eq.coefa1}
	&a_1^{++}=\mathcal{M}_{{qB}}^{\mu\nu}\hat{L}_{1\mu}\hat{L}_{2\nu},\ \
	a_2^{++}=\mathcal{M}_{{qB}}^{\mu\nu}\hat{L}_{1\mu}^{*}\hat{L}_{2\nu}^{*}\\
    &\mbox{and} \ \
	a_4^{+-}=\mathcal{M}_{{qB}}^{\mu\nu}\dfrac{1}{\sqrt{2}}\left(\hat{L}_{1\mu}\hat{L}_{2\nu}^{*}+\hat{L}_{1\mu}^{*}\hat{L}_{2\nu}\right).
 \end{split}
\end{equation}
Then, writing the amplitude in terms of an orthogonal tensor structure decomposition helps to compute the physical observables in a simpler way. Recently, a new methodology was developed to compute the amplitude shown in Fig.\ref{figuratriagulo}, for an arbitrary magnetic field strength~\cite{JaberSanchez}. 

The above ideas are general in the sense that the asymptotic states, expressed in an useful form, help to extract in a simple way the magnetic field effects on more involved processes, as happens, for example, in the work presented in Ref.~\cite{ScoccolaRev} which deals with asymptotic charged states. There, the effect of an homogeneous external magnetic field on the charged pion decay width to leptons is analyzed $\left(\pi^{-}\longrightarrow l^{-}\bar{\nu}_{l}\right)$.

In the vacuum, the total width is dominated by the muonic channel, for which the branching ratio reaches about 99.99\%, due to ``helicity suppression''.

The transition matrix of the processes can be obtained from the interaction of the correspondent hadronic current with the asymptotic spinors states of leptons. Here, the hadronic and leptonic sectors are described with their correspondent states in presence of an uniform magnetic field in the Ritus formalism. In particular, the magnetic field increases the number of independent tensor structures and three more form factors appear in the hadronic current. In order to obtain an expression for the decay width, the pion is supposed to be in the LLL and the integration over the phase space of the product particles is performed.

Within the context of the NJL model, the values of the four form factors are taken from~\cite{ScoccolaNJL},  in order to obtain numerical results. The results indicate a large enhancement of the decay width with respect to the vacuum case, for both muonic and electronic channels and also indicate that there is no ``helicity suppression'', \textit{i.e.}, the decay to an electron is facilitated by the magnetic field. In addition, it is observed that the angular distribution of the outgoing antineutrinos is expected to be suppressed in the direction of the magnetic field.

The probability of pair creation ($e^- e^+$) by a photon in a strong magnetic field was (re)analyzed in \cite{Baier}, again through the imaginary part of the self-energy, expanding the energy interval for photons and allowing for off-shell processes. After a historical outlook of the study of the pair creation process, the exact probability is exhibited, with an intrincate dependence on the magnetic field: $B$ multiplied by the exponential ($\exp{\left(-p_{\perp}^2/{B}\right)}$) and keeping the sum over Landau levels. Their graphs for the decay amplitude show the typical spike structure associated with the different Landau levels, that is afterwards smoothed by averaging over an interval of photon energy distribution. They finally compare their outcome with previous on-shell results (e.g. Tsai and Erber) for weak and strong field approximations, finding in general a good agreement.

In \cite{Karbstein}, the photon polarization tensor in a homogeneous magnetic (or electric) field was also reexamined, both on- and off-the-light-cone dynamics, paying particular attention to the systematization  of the different approximation schemes for well-constrained parameter regimes. Taking the photon self-energy imaginary part, a comparison with the results of \cite{TSAI.1,TSAI.2} and \cite{Baier} was performed, obtaining a very good coincidence with the last one and agreement with the former at large $p_\perp$ values.

\section{Conclusions}

 The aim of our review was to put together and compare different results and approaches that can be found in the literature about the effects of magnetic fields on particle interactions. The importance of the magnetic field contribution comes from the fact that they are widespread in the Universe and in high-energy physics experiments. 

The literature in this subject is so vast that the present work is far from being exhaustive, we mainly focused on some seminal papers and on recent literature, both from our working group and from other research groups.
Facing the situation that the results presented in the literature can be very different between one situation and another, our interest fell on finding a scheme that could be somehow predictive, once all the physical ingredients are taken into account. We did not fully succeeded since we still face some discrepant results, but we arrived to some preliminary conclusions. 

First of all, since the analytical calculations require of an approximation approach, it is clear that it is very important to have a good control on the validity of the different approximation schemes that hold for different well-constrained parameter regimes, in particular for the transferred momentum regime.
On another hand, there is in general an agreement that strong magnetic fields favor the interaction, with a linear dependence of the decay width on the magnetic field strength, that emerges from processes that involve charged fermions, since these are confined into LLL in this regime.

In contrast, for weak magnetic fields, the situation is far from clear: the process is favored in some cases, inhibited in some others, or it can depend on the kinematical regime of the involved particles. Besides the transferred momentum, another important ingredient for these differences could be the spin of the particles involved in the interaction. Although the different results can be traced back to the different analytical structures of the propagators of the particles involved in the interaction, the physical motivation for such behaviors still needs more discussion.

If the decrease of the decay rate with increasing magnetic field, that emerges in some situations at low-momentum, can be due to the phase space reduction, for high momenta the underlying physical explanation is more puzzling. A possible explanation that has been offered is: if we imagine that the (composite) decaying particle were formed by a pair of charged particles, then the Lorentz force would act in opposite directions on each forming particle, promoting the decay process.

Finally, regarding another important ingredient -the spin- in order to shed some light on its r\^ole,  the processes of decay of a scalar to two charged scalars and to two charged fermions were compared, in the high momentum regime. The magnetic field impact was observed to be more pronounced for fermions than for scalars. This result may indicate that the spin-magnetic field
interaction increases the pair production in the fermion case.

There is still much work to be done in this field to systematically explore the contribution of all the possible physical ingredients in all the possible regimes.

\section*{Acknowledgments}

Support for this work has been received in part from DGAPA-UNAM under grant numbers PAPIIT-IN117023 and PAPIIT-IN108123. G.P. thanks the hospitality of Facultad de Ciencias, UNAM, during a sabbatical stay, where this work was completed. J.J.U. acknowledges the support received by CONACyT fellowship No. 780045.

\bibliography{Template}

\end{document}